\documentclass{article}

\usepackage[margin=1in]{geometry}
\usepackage{amsfonts, amsmath}
\usepackage{dsfont}
\usepackage{graphicx}
\usepackage{url}
\usepackage{hyperref}

\newcommand{\footremember}[2]{%
   \footnote{#2}
    \newcounter{#1}
    \setcounter{#1}{\value{footnote}}%
}
\newcommand{\footrecall}[1]{%
    \footnotemark[\value{#1}]%
}

\usepackage{fancyhdr}
\title{Inferring medication adherence from time-varying health measures}
\author{%
Kristen B. Hunter\footremember{Harvard}{Department of Statistics, Harvard University}, %
Mark E. Glickman\footrecall{Harvard}, %
Luis F. Campos\footnote{Etsy, Inc.}%
}
\date{\today}

\begin{document}

\maketitle

\section*{Abstract}
Medication adherence is a problem of widespread concern in clinical care. 
Poor adherence is a particular problem for patients with chronic diseases requiring long-term medication because poor adherence can result in less successful treatment outcomes and even preventable deaths. 
Existing methods to collect information about patient adherence are resource-intensive or do not successfully detect low-adherers with high accuracy.
Acknowledging that health measures recorded at clinic visits are more reliably recorded
than a patient's adherence, we have developed an approach to infer medication adherence rates based on longitudinally recorded health measures that are likely impacted by time-varying adherence behaviors. 
Our framework permits the inclusion of baseline health characteristics and socio-demographic data. 
We employ a modular inferential approach.
First, we fit a two-component model on a training set of patients who have detailed adherence data obtained from electronic medication monitoring.
One model component predicts adherence behaviors only from baseline health and socio-demographic information, and the other predicts longitudinal health measures given the adherence and baseline health measures.
Posterior draws of relevant model parameters are simulated from this model using Markov chain Monte Carlo methods.
Second, we develop an approach to infer medication adherence from the time-varying health measures 
using a Sequential Monte Carlo algorithm applied to a new set of patients for whom no adherence data are available. 
We apply and evaluate the method on a cohort of hypertensive patients, using baseline health comorbidities, socio-demographic measures, and blood pressure measured over time to infer patients' adherence to antihypertensive medication.

\textbf{Keywords:} Medication adherence, state-space models, Sequential Monte Carlo, hypertension.

%%%%%%%%%%%%%%%%%%%%%%%%%%%%%%%%%%%%%%%%%%%%%%%%%%%%%%%%%%%%%%%%%%%%%%%%%%%%%%%%%%%%%%%%%%
% Introduction and motivation
%%%%%%%%%%%%%%%%%%%%%%%%%%%%%%%%%%%%%%%%%%%%%%%%%%%%%%%%%%%%%%%%%%%%%%%%%%%%%%%%%%%%%%%%%%

\section{Introduction}
\label{sec:intro}

Patients' non-adherence to their prescribed medication is a serious obstacle to successful medication therapy and a widespread problem in clinical care \cite{Vitolins00, Johnson02, Lam15, who03}.
Not only can a lack of medication adherence lower the impact of successful treatment, but poor adherence can ultimately increase medical costs and health care utilization for treating worsening health conditions \cite{Hughes07}.
In more serious disease settings, poor adherence can result in preventable deaths \cite{Farley10}.
As an example of the extent of the problem, the WHO \cite{who03} reported that adherence for long-term medications treating chronic diseases in developed countries averages only 50\%.
With poor adherence being a substantial barrier to effective medical treatment, it is imperative to find ways to improve medication adherence and thereby prevent needless adverse medical consequences and avoid unnecessary drainage of resources to the health system.

Providers and patients are likely more empowered to make informed decisions if they have accurate information about medication adherence.
Accurate assessment of adherence behaviors can help guide a discussion between patient and provider as to the underlying causes of poor adherence with the goal of improving adherence rates.
Providers can then educate patients on adherence best practices based on self-efficacy behaviors.

Current methods to collect medication adherence information are generally not practical or accurate enough to be useful in clinical settings.
Most existing methods involve collecting additional data from patients or pharmacies \cite{Lam15}.
Patient self-reporting is the simplest way to collect adherence information, but it has been shown
to be an unreliable method, particularly for detecting low-adherers \cite{Hamilton03, Shalansky04, Svarstad99, Zeller08}. One reason for this unreliability is that patient memory about adherence is often poor, and typically inflated.
For example, Waterhouse \cite{Waterhouse93} found that patients' self-reported adherence levels were three times higher than adherence levels recorded by electronic monitoring.
Shalansky et al. \cite{Shalansky04} fit a logistic regression of medication adherence as a function of the Morisky adherence score \cite{Morisky86}, a self-assessment tool, and found it to be a significant predictor.
The model, however, had low sensitivity.
Although 13\% of patients were considered nonadherent, only 3\% had a Morisky score corresponding to a high risk of nonadherent behavior.
Medication adherence levels have also been measured through electronic bottle caps, pharmacy refill rates, and counts of remaining pills from bottles returned by patients
\cite{KrouselWood04}.
Though more accurate than self reporting, these methods can be resource-intensive, and thus difficult to implement in clinical practice.
Methods such as pharmacy refill rates and pill counts also do not guarantee accuracy.
For example, although a patient may regularly refill their prescription, that does not mean they are regularly taking the medication.
Also, pill counts rely on patients returning medication bottles and not disposing of any pills.

An alternative approach involves modeling adherence behavior.
However, most previous studies have found it challenging to explain much of the variation in adherence.
Balkrishnan et al.\cite{Balkrishnan03} modeled medication dispensation rates for type 2 diabetes patients using a random effects generalized least squares model and found that high comorbidity severity and an ER visit in the last year were significant predictors of low adherence.
The overall model reported an $R^2$ statistic of $0.3$.
Wu et al. \cite{Wu08} modeled medication adherence for patients with heart failure using a multiple linear regression.
The authors included a rich set of variables in their models, including self-reported survey responses, patient-related data, socioeconomic information, and condition, treatment, and health system information, but were only able to explain $11-21\%$ of the variation in adherence.
Finally, Yue et al. \cite{Yue15} conducted univariate analyses on the association between a variety of patient covariates and antihypertensive medication adherence level and found only age to have a significant association with adherence.
These studies highlight the difficulty of modeling adherence behaviors on baseline covariates alone.

In this paper, we develop a novel approach to infer medication adherence rates from commonly-collected clinical data.
Our method infers a patient's recent adherence behavior using two sources: health measures recorded over time that are likely to be directly impacted by differential adherence, and baseline covariates including health characteristics and sociodemographic data.
Our approach to infer adherence has advantages over existing practices to infer adherence behavior.
We provide a data-driven estimate of adherence that may be more reliable than self-reporting.
Additionally, we exploit data that are already collected in routine clinical care, so no additional data collection is needed.
Finally, we use time-varying health measures, which are not often used in adherence models.
Our model is trained on a set of patients with detailed adherence information collected using electronic monitoring.
The training step requires specialized equipment, in our case electronic bottle caps, which may be more resource-intensive than methods such as tracking pharmacy refill rates.
However, once the model has been trained on a sufficiently large and representative sample of patients, it can be applied to patients in the same population using only routinely-collected clinical data.
The resulting computation to estimate unobserved adherence could serve as the foundation of a clinical decision tool used by patients and providers in real time.

We construct a modular procedure to infer adherence values for patients with unknown adherence behavior.
The first step involves a cohort of patients designated as the training set.
The training set is used to fit a model of observed adherence as a function of baseline covariates.
The posterior distribution based on this model determines the prior probability of adherence for a new patient, before any health measures are observed.
Separately, a state space model (SSM) relating adherence and health measures is calibrated using the training set.
We use Markov chain Monte Carlo (MCMC) simulation to fit both the adherence model and health measures SSM on the training data.
The second step uses a cohort of patients designated as the test set.
With the posterior distributions from the adherence and health measures models, we can infer longitudinal patterns of adherence for a new patient, given a time series of health measures and their baseline covariates.
For the purpose of inferring adherence, we use the same state space model employed above, and we view the task of estimating adherence as a ``smoothing problem,'' in the parlance of state space models.
Smoothing problems can be generically tackled with sequential Monte Carlo methods \cite{Doucet08,Kantas15}, which is the approach we take in our framework.

We applied our approach to a cohort of hypertensive patients, all of whom were prescribed antihypertensive medications.
For each patient, we obtained daily indicators of whether they were adherent to their medication using electronic monitoring, irregularly-measured readings of systolic and diastolic blood pressure from routine clinical care, and baseline covariates and comorbidities.
The above procedure was then used to generate predictions of adherence for patients in the test set.
We found that the posterior intervals for estimated average adherence displayed good coverage properties, and the interval lengths appear narrow enough to be of practical use in a clinical setting.

This paper proceeds as follows.
Section~\ref{sec:model} outlines the different models, including a logistic regression model for adherence and a Normal linear state space model for health measures.
We also discuss obtaining posterior inferences from these models using MCMC.
Section~\ref{sec:inference} develops the method for inferring a patient's unobserved medication adherence using SMC based on the fit of the SSM.
The modular nature of the inferential procedure is introduced.
We also include details of the particle Gibbs with ancestor sampling \cite{Lindsten14} algorithm used in our approach.
Section~\ref{sec:hyp} demonstrates our method applied to a study examining the relationship between antihypertensive medication adherence and blood pressure measures.
We determine interval estimates of adherence for a withheld patient sample.
Finally, Section~\ref{sec:discussion} concludes.

%%%%%%%%%%%%%%%%%%%%%%%%%%%%%%%%%%%%%%%%%%%%%%%%%%%%%%%%%%%%%%%%%%%%%%%%%%%%%%%%%%%%%%%%%%
% A joint model for adherence and health outcomes
%%%%%%%%%%%%%%%%%%%%%%%%%%%%%%%%%%%%%%%%%%%%%%%%%%%%%%%%%%%%%%%%%%%%%%%%%%%%%%%%%%%%%%%%%%

\section{A joint model for adherence and health measures}
\label{sec:model}

We specify our model for medication adherence and health measures in two parts. Section \ref{sec:adherence_model} introduces a random intercept logistic regression model for adherence as a function of baseline covariates, with a random effect for each patient. 
Section \ref{sec:outcomes_model} introduces a Normal linear state space model for observed health measures as a function of adherence and baseline covariates. 

%%%%%%%%%%%%%%%%%%%%%%%%%%%%%%%%%%%%%%%%%%%%%%%%%%%%%%%%%%%%%%%%%%%%%%%%%%%%%%%%%%%%%%%%%%
% Adherence model
%%%%%%%%%%%%%%%%%%%%%%%%%%%%%%%%%%%%%%%%%%%%%%%%%%%%%%%%%%%%%%%%%%%%%%%%%%%%%%%%%%%%%%%%%%

\subsection{Adherence model}
\label{sec:adherence_model}

For $i=1,\ldots,N$, where $N$ is the number of patients in the cohort, 
let $T_i$ be the number of consecutive days patient $i$ was in the study.
Let $c_{it}$, for $t=1,\ldots,T_i$, be an indicator for medication adherence, where $c_{it} = 1$ if patient $i$ took their medication on day $t$ and $c_{it} = -1$ if patient $i$ did not take their medication on day $t$.
We assume a linear predictor $\delta_i + \boldsymbol{\lambda}^\top\boldsymbol{x}_i$, where $\delta_i$ is the random effect for patient $i$ with $\delta_i \sim N(0, \sigma_{\delta})$, $\boldsymbol{x}_i$ is a vector of $p$ covariates, and $\boldsymbol{\lambda}$ is the vector of covariate coefficients. Then the adherence model is
\begin{align}\label{eqn:c}
p(c_{it}\mid \boldsymbol{x}_i, \delta_i, \boldsymbol{\lambda})
&= \left(\frac{\exp\left(\delta_i + \boldsymbol{\lambda}^\top\boldsymbol{x}_i \right)}{1 + \exp\left(\delta_i + \boldsymbol{\lambda}^\top \boldsymbol{x}_i\right)}\right)^{\frac{c_{it} + 1}{2}}\left(\frac{1}{1 + \exp\left(\delta_i + \boldsymbol{\lambda}^\top\boldsymbol{x}_i\right)}\right)^{1 - \frac{c_{it} + 1}{2}}.
\end{align}
We will later use the notation $\theta_a$ to refer to the model parameters $\theta_a = \{\boldsymbol{\lambda}, \sigma_\delta\}$, common to all patients, for our adherence model.

This model for adherence assumes that a patient's adherence is not time-varying.
An alternative approach would be to model the longitudinal variation in daily adherence observations.
However, accurate modeling of time variation in medication-taking is difficult, in part because patients vary considerably in the ways in which they do not adhere.
For example, some patients are consistently adherent, some may become less adherent over time, some are adherent during the weekdays but not on weekends, and so on.
Thus, even longitudinal studies of adherence, such as the work of Vrijens et al. \cite{Vrijens08}, often focus on modeling the time until a patient 
has become entirely non-adherent.
We expect modeling daily variation in medication adherence to be an unrealistic goal, so instead we aim to infer and summarize average adherence rates over time intervals that may be clinically meaningful.
Acknowledging these limitations, we adopt a simple random effects logistic regression to capture the variation in adherence across patients.

Our adherence model is a special case of a generalized linear mixed model.
Bayesian inference for such models commonly uses numerical methods via posterior simulation, and has been developed in the context of MCMC sampling in Chib and Carlin \cite{Chib99}.
The choice of a prior distribution is case-specific, and is discussed for our application in Section~\ref{sec:posterior}.
We perform posterior simulation using Hamiltonian Monte Carlo \cite{Duane87}, which can be implemented using the software package STAN \cite{Carpenter17}.

%%%%%%%%%%%%%%%%%%%%%%%%%%%%%%%%%%%%%%%%%%%%%%%%%%%%%%%%%%%%%%%%%%%%%%%%%%%%%%%%%%%%%%%%%%
% Health outcomes model
%%%%%%%%%%%%%%%%%%%%%%%%%%%%%%%%%%%%%%%%%%%%%%%%%%%%%%%%%%%%%%%%%%%%%%%%%%%%%%%%%%%%%%%%%%

\subsection{Health measures model}
\label{sec:outcomes_model}

We assume that health measures follow a Normal linear state space model.
State-space models are a flexible way to model time series outcomes in which some variables are assumed to follow a latent stochastic process, and observed outcomes are distributed according to a model conditional on the parameters specific to that time point \cite{Cappe05, Petris09, West97, Durbin01}.
A summary of the proposed state-space model follows.
Further details of this model are described and developed in Campos et al. \cite{Campos20} for general settings involving the time-varying effects of medication adherence on health measures.

Our model assumes a $K$-dimensional vector of outcome measures recorded over time and at possibly irregular intervals.
Let $y_{itk}$ be the observed measurement for patient $i$ at time $t$ for health outcome $k=1,\ldots,K$.
We denote the collection over an index using dot notation.
For example, $\boldsymbol{y}_{it\cdot}$ denotes the $K$-vector of health outcomes $(y_{it1} , y_{it2}, \ldots, y_{itK} )^\top$, and $\boldsymbol{y}_{i\cdot\cdot}$ further collects over the time points to form a $K \times T_i$ matrix.
Similarly, $\boldsymbol{c}_{i\cdot}$ is the $T_i$-vector of adherence values for unit $i$.

We assume that $\boldsymbol{y}_{it\cdot}$ is multivariate Normal conditional on model parameters,
\begin{align}\label{eqn:y}
\boldsymbol{y}_{it\cdot} &= \boldsymbol{\beta}\boldsymbol{x}_i + \boldsymbol{\alpha}_{it\cdot} + \boldsymbol{\varepsilon}_{it\cdot},
\end{align}
where $\boldsymbol{\beta}$ is a $K \times p$ coefficient matrix, the $\boldsymbol{\alpha}_{it\cdot}$ are $K$-dimensional latent time-varying effects that depend on adherence, and $\boldsymbol{\varepsilon}_{it\cdot}$ is a $K$-vector of zero-mean, normally distributed measurement errors with covariance matrix $\boldsymbol{\Sigma}_{\varepsilon}$. 

We assume that the $\boldsymbol{\alpha}_{it\cdot}$ are modeled as an AR(1) process according to
\begin{align}\label{eqn:alpha}
\boldsymbol{\alpha}_{it\cdot} &= \boldsymbol{\rho} \boldsymbol{\alpha}_{i,t-1,\cdot} + \boldsymbol{\phi} c_{it} + \boldsymbol{\nu}_{it\cdot},
\end{align}
where $\boldsymbol{\rho} = \text{diag}(\rho_1, \rho_2, \dots \rho_K)$ is a $K \times K$ matrix specifying the auto-correlation between states for each health outcome, the $K$-vector $\boldsymbol{\phi}$ contains the coefficients to the adherence indicator, and $\boldsymbol{\nu}_{it\cdot}$ is a $K$ vector of independent, zero-mean, normally distributed innovations for the latent processes with diagonal covariance matrix $\boldsymbol{\Sigma}_\nu = {\text{diag}}(\sigma^2_{\nu 1}, \sigma^2_{\nu 1}, ..., \sigma^2_{\nu K})$.
We assume $\alpha_{i1} \sim \mathcal{N}_K(0, \boldsymbol{\Sigma}_0)$, where $\boldsymbol{\Sigma}_0 = \text{diag}(\sigma_{01}^2, \sigma_{02}^2, \dots, \sigma_{0K}^2)$. We will later use the notation $\theta_h$ to refer to the model parameters common to all patients for the health measures model, $\theta_h = \{\boldsymbol{\beta}, \boldsymbol{\rho}, \boldsymbol{\phi},\Sigma_\varepsilon, \Sigma_\nu, \boldsymbol{\Sigma}_0\}$.

We emphasize that generally the number of observed health measures is much smaller than the number of observed adherence values.
The health measures $\boldsymbol{y}_{it\cdot}$ are only observed at irregularly-spaced intervals, such as during clinic visits by a patient.
In contrast, the adherence values $c_{it}$ are typically measured regularly, often daily, through electronic monitors.

The parameters $\theta_h$ for the health measures model can be inferred using Bayesian computational strategies described in Campos et al. \cite{Campos20}.
The approach acknowledges the irregularly-spaced health measures, and efficiently performs inference by factoring the posterior density into the density of the non-dynamic parameters that integrates out the time-varying variables $\boldsymbol{\alpha}_{it\cdot}$, and the conditional density of the time-varying variables given the non-dynamic parameters.
In particular, the distribution of 
$\boldsymbol{y}_{i\cdot\cdot}$ given $\theta_h$ and $\boldsymbol{c}_{i\cdot}$ is Normal and can be evaluated with the Kalman filter.
Thus the likelihood function is tractable and a variety of MCMC algorithms can be implemented to approximate the posterior distribution of $\theta_h$.
Further, for any $\theta_h$ and given $\boldsymbol{y}_{i\cdot\cdot}$ and $\boldsymbol{c}_{i\cdot}$ the latent variables $\boldsymbol{\alpha}_{i\cdot\cdot}$ can be retrieved by performing a task
called ``smoothing'' in the context of state space models.
Here it can be done exactly using the Kalman smoother.
As with the adherence model, the prior distributions are chosen specific to the applied context; details for our application are in Section~\ref{sec:posterior}.
Although we assume Normality, state-space models are flexible and can easily be extended to non-Normal distributions, although a different computational approach would then be needed.

%%%%%%%%%%%%%%%%%%%%%%%%%%%%%%%%%%%%%%%%%%%%%%%%%%%%%%%%%%%%%%%%%%%%%%%%%%%%%%%%%%%%%%%%%%
% Predictive inference
%%%%%%%%%%%%%%%%%%%%%%%%%%%%%%%%%%%%%%%%%%%%%%%%%%%%%%%%%%%%%%%%%%%%%%%%%%%%%%%%%%%%%%%%%%

\section{Predictive inference for unobserved adherence}
\label{sec:inference} 

\subsection{Assembling models to predict adherence}

Our inferential goal parallels a clinical setting in which we want to infer past adherence behavior for a new patient with no previously observed adherence measures, but with baseline covariates and health measures available.
Thus, we are ultimately interested in conducting inference on the set of adherence values $\boldsymbol{c}_{i\cdot} = (c_{i1} , \ldots , c_{iT_i} )^\top$,
or a summary of these values, for new patient $i$ for the $T_i$ days over which patient $i$ is monitored.

Before conducting inference, we randomly split patients into a training set, containing fully observed adherence and health measures, and a test set, where we mask the adherence and only use health measures.
This split allows us to evaluate the performance of the procedure on a held-out test set.
When fitting the model for use in clinical practice, we would include all available patients in the training set, and the `test' set would be future patients for whom we want to predict adherence.
We chose to fit our model using a training set and then produce predictions on a test set of patients because it allows us to evaluate performance in a setting that arguably most closely mirrors clinical practice.
Ideally, the goal is to apply this same approach in clinical practice; the model would be trained on a set of patients, then applied to new patients using only available health information from their health records and from repeated blood pressure levels (or, more generally, repeated health measures) obtained at clinic visits over time.
Another method for evaluating the performance of our approach would be to fit the state-space model on the entire cohort, but leave out a final period of days of medication-taking information for each patient, and infer adherence over this later period.
This approach would take advantage of adherence data within an individual patient to make stronger inferences about unknown adherence rates in the later periods.
However, such an approach would not provide as robust a measure of performance on a held-out test set, which is our metric of interest.

After splitting the patients into training and test sets, we then perform a modular, multi-step procedure.
First, we obtain draws of model parameters using the training set via posterior sampling.
This procedure is conducted in two separate steps: the adherence model parameters are inferred separately from the health measures model parameters.
Next, we produce draws of estimated adherence for the test patients given the observed health measures and model parameters.
One advantage of this procedure is that we obtain draws of the full set of daily estimated adherence for each patient, rather than a summary of adherence.
Although we focus on average adherence as a way to summarize the posterior draws, the more detailed information would also be available for further exploration.

We now provide further details and justification of the algorithm.
The training set, consisting of $n_{train}$ patients, contains adherence data, baseline covariates, and time-varying health measures.
We assume baseline covariates are not missing for any patients.
Adherence data is assumed to be observed on a daily basis.
In contrast, health measures may only be observed irregularly.
To simplify notation, the set of training data is denoted as $\{\boldsymbol{y}^\prime, \boldsymbol{x}^\prime, \boldsymbol{c}^\prime\}.$
The matrix $\boldsymbol{x}^\prime$ is a $p \times n_{train}$ matrix of covariate values.
The array $\boldsymbol{c}^\prime$ is a 2-D ragged array of adherence indicators with $n_{train}$ columns, where column $j$ (denoted $\boldsymbol{c}_{j\cdot}$) consists of $T_j$ values for $j = 1,\ldots, n_{train}$.
The array $\boldsymbol{y}^\prime$ is a 3-D ragged array of health measure values, where matrix $j$ (denoted $\boldsymbol{y}_{j\cdot\cdot}$) consists of $T_j \times K$ values, some of which are potentially missing, for $j = 1,\ldots, n_{train}$.

We want to infer a summary of the adherence values for each new patient $i$ in the test set conditional on that patient's covariates and health measures, in addition to the data from the $n_{train}$ patients in the training set.
While the procedure below is described for a single test patient, we may consider a total of $n_{test}$ new patients for whom the procedure will apply, so that $i=1,\ldots, n_{test}$.
For each patient $i$ in the test set, we denote $\boldsymbol{c}_{i\cdot}^\star$ as the fully unobserved vector of adherence values, and $\boldsymbol{y}^\star_{i\cdot\cdot}$ as the $K \times T_i$ matrix of potentially observed health measures.
Inference can be accomplished via posterior predictive sampling from $p(\boldsymbol{c}_{i\cdot}^\star \mid \boldsymbol{y}^\star_{i\cdot\cdot}, \boldsymbol{x}_i^\star, \boldsymbol{y}^\prime, \boldsymbol{x}^\prime, \boldsymbol{c}^\prime)$, for $i = 1, \dots, n_{test}$.
In what follows, we suppress the conditioning on $\boldsymbol{x}_i^\star$ and $\boldsymbol{x}^\prime$ for improved readability, and focus on
the approximation of $p(\boldsymbol{c}_{i\cdot}^\star \mid \boldsymbol{y}^\star_{i\cdot\cdot}, \boldsymbol{y}^\prime, \boldsymbol{c}^\prime)$.

Introducing the parameters $\theta_a$ and $\theta_h$, we write
\begin{align} \label{eq:distribC}
p(\boldsymbol{c}_{i\cdot}^\star \mid \boldsymbol{y}^\star_{i\cdot\cdot}, \boldsymbol{y}^\prime, \boldsymbol{c}^\prime)
&= 
\int p(\boldsymbol{c}_{i\cdot}^\star \mid \boldsymbol{y}^\star_{i\cdot\cdot}, 
\theta_a, \theta_h)
p(\text{d}\theta_a, \text{d}\theta_h
\mid \boldsymbol{y}^\star_{i\cdot\cdot}, \boldsymbol{y}^\prime, \boldsymbol{c}^\prime).
\end{align}
The above relies on $\boldsymbol{c}_{i\cdot}^\star \mid \boldsymbol{y}^\star_{i\cdot\cdot}, \boldsymbol{y}^\prime, \boldsymbol{c}^\prime,\theta_a,\theta_h$ being independent of $\boldsymbol{y}^\prime, \boldsymbol{c}^\prime$, which follows from the model specifications.

Next we make an approximation: we replace 
$p(\text{d}\theta_a, \text{d}\theta_h \mid \boldsymbol{y}^\star_{i\cdot\cdot}, \boldsymbol{y}^\prime, \boldsymbol{c}^\prime)$
by $p(\text{d}\theta_a, \text{d}\theta_h \mid \boldsymbol{y}^\prime, \boldsymbol{c}^\prime)$.
The information provided by the health measures, $\boldsymbol{y}^\star_{i\cdot\cdot}$, is limited given the information already provided by the training data, $(\boldsymbol{y}^\prime, \boldsymbol{c}^\prime)$, particularly given that the outcomes are not paired with adherence information.
The additional health measures from the test set would only influence the posterior distribution of the intercepts and covariate coefficients $\boldsymbol{\beta}$, and we assume that the training data is sufficient for making inferences on these parameters.
Thus, although we trade off a minimal loss of information, the computational benefit of not refitting the model using the test set is advantageous in our framework.
Thanks to this approximation, we can consider drawing values of $\theta_a,\theta_h$ from $p(\text{d}\theta_a, \text{d}\theta_h
\mid \boldsymbol{y}^\prime, \boldsymbol{c}^\prime)
=
p(\text{d}\theta_a
\mid \boldsymbol{c}^\prime)
p(\text{d}\theta_h
\mid \boldsymbol{y}^\prime, \boldsymbol{c}^\prime)$;
i.e., we draw $\theta_a$ from the posterior of the adherence model and $\theta_h$ from the posterior of the health measures model.
For each pair $\theta_a,\theta_h$, we can then draw samples from $p(\boldsymbol{c}_{i\cdot}^\star \mid \boldsymbol{y}^\star_{i\cdot\cdot}, \theta_a, \theta_h)$, which is a smoothing task described below.
The different steps above can be summarized by the replacement of \eqref{eq:distribC} by
\begin{align} \label{eq:distribCsimplified}
p(\boldsymbol{c}_{i\cdot}^\star \mid \boldsymbol{y}^\star_{i\cdot\cdot}, \boldsymbol{y}^\prime, \boldsymbol{c}^\prime)
\approx
\int 
\underbrace{p(\boldsymbol{c}_{i\cdot}^\star \mid \boldsymbol{y}^\star_{i\cdot\cdot}, 
\theta_a, \theta_h)}_{\text{smoothing model}}
\underbrace{p(\text{d}\theta_a
\mid \boldsymbol{c}^\prime)}_{\text{adherence model    }}
\underbrace{p(\text{d}\theta_h
\mid \boldsymbol{y}^\prime, \boldsymbol{c}^\prime)}_{\text{health measures model}}.
\end{align}

We now move on to the approximation of $p(\boldsymbol{c}_{i\cdot}^\star \mid \boldsymbol{y}^\star_{i\cdot\cdot}, \theta_a, \theta_h)$. Using Bayes rule, this distribution is proportional to $p(\boldsymbol{c}_{i\cdot}^\star|\theta_a)
p(\boldsymbol{y}^\star_{i\cdot\cdot}, 
\mid \boldsymbol{c}_{i\cdot}^\star,
\theta_a, \theta_h)$.
The latter factor does not depend on $\theta_a$, and is a Normal linear 
state space model as described in Section~\ref{sec:outcomes_model}, with latent process on $\boldsymbol{\alpha}_{i\cdot}$.
The former term gives a prior distribution of 
$p(\boldsymbol{c}_{i\cdot}^\star|\theta_a)$ in the form of a product
of Bernoulli distributions, according to the adherence model.
Overall we can thus view $(\boldsymbol{c}_{i\cdot},\boldsymbol{\alpha}_{i\cdot})$ as a latent process and $\boldsymbol{y}_{i\cdot\cdot}$ as a series of observations.
This augmented state space model is thus not a Normal linear model, because the latent process is made of both discrete and continuous variables.
Specialized algorithms have been devised for this type of model \cite[e.g.]{andrieu2002rao}, but we will employ instead generic particle methods that can more easily accommodate alternative specifications of the model.

\begin{figure}
\centering
\includegraphics[width = 4in]{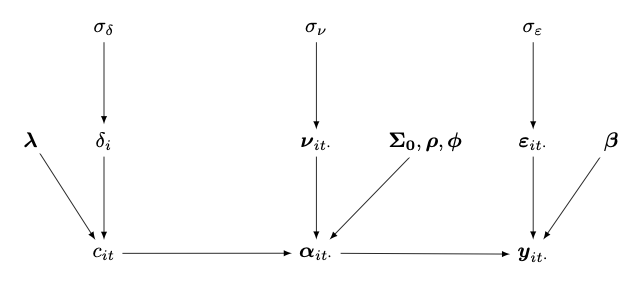}
\caption{\label{fig:diag}Structure of the model parameters.}
\end{figure}

To summarize, our procedure follows naturally from the final expression in~\eqref{eq:distribCsimplified}.
First, collecting terms into $\boldsymbol{\theta} = \{\theta_a, \theta_h\}$, we obtain draws from the distribution $p(\boldsymbol{\theta} \mid \boldsymbol{y}^\prime, \boldsymbol{c}^\prime)$.
Inference can be conducted independently on the adherence model and the health measures model.
This approach is valid because the posterior distribution of $\boldsymbol{\theta}$ can be factored into two parts, one containing the components of the adherence model $\theta_a$, and one containing the components of the health measures model $\theta_h$, so the distributions can be estimated separately.
Figure \ref{fig:diag} summarizes the structure of the model parameters.
The procedures for sampling $\theta_a$ and $\theta_h$ are discussed in Sections~\ref{sec:adherence_model} and \ref{sec:outcomes_model}.

Second, conditional on each sampled value of $\boldsymbol{\theta}$, we sample from the distribution $p\left(\boldsymbol{c}_{i\cdot}^\star \mid \boldsymbol{y}_{i\cdot\cdot}^\star, \boldsymbol{\theta} \right)$. Specifically, we will employ sequential Monte Carlo methods as described in Section~\ref{sec:smc}.
Finally, we collect all sampled adherence vectors $\boldsymbol{c}_{i\cdot}^\star$, define our adherence quantity of interest, and summarize the Monte Carlo distribution of this quantity.
This procedure mirrors a setting in which we have previously trained a model based on a set of patients, and now want to infer adherence behavior for new patients.

As mentioned above, because the resulting state space model is not Normal, we cannot use Kalman smoothing and we propose to use a generic sequential Monte Carlo (SMC) approach to sample from 
$p(\boldsymbol{c}_{i\cdot}^\star \mid \boldsymbol{y}_{i\cdot\cdot}^\star, \boldsymbol{\theta})$.
Instead of SMC sampling, a more direct approach would be to use importance sampling, by relying on the expression 
\begin{align} \label{eq:bayes1}
  p(\boldsymbol{c}_{i\cdot}^\star \mid \boldsymbol{y}_{i\cdot\cdot}^\star, \boldsymbol{\theta})
  \propto
  p\left(\boldsymbol{c}_{i\cdot}^\star \mid \boldsymbol{\theta} \right)
  p\left(\boldsymbol{y}_{i\cdot\cdot}^\star \mid \boldsymbol{c}_{i\cdot}^\star , \boldsymbol{\theta}\right) .
\end{align}
This equation suggests that given $\boldsymbol{\theta}$ we could sample from $p\left(\boldsymbol{c}_{i\cdot}^\star \mid \boldsymbol{\theta} \right)$, which is a product of Bernoulli distributions,  and then re-weight the samples proportional to $p\left(\boldsymbol{y}_{i\cdot\cdot}^\star \mid \boldsymbol{c}_{i\cdot}^\star , \boldsymbol{\theta}\right)$, evaluated, for example, using the Kalman filter, to obtain weighted draws approximating $p(\boldsymbol{c}_{i\cdot}^\star \mid \boldsymbol{y}_{i\cdot\cdot}^\star, \boldsymbol{\theta})$.
This direct approach can be quite inefficient because adherence vectors sampled from $p\left(\boldsymbol{c}_{i\cdot}^\star \mid \boldsymbol{\theta} \right)$ are likely to fall in regions of nearly zero posterior mass.
Sequential Monte Carlo is a more efficient approach that generates draws of $\boldsymbol{c}_{i\cdot}^\star$ using the information contained in the observations $\boldsymbol{y}_{i\cdot}$, by implementing forward (and/or backward) sweeps over the observations.

%%%%%%%%%%%%%%%%%%%%%%%%%%%%%%%%%%%%%%%%%%%%%%%%%%%%%%%%%%%%%%%%%%%%%%%%%%%%%%%%%%%%%%%%%%
% Overview of Sequential Monte Carlo methods
%%%%%%%%%%%%%%%%%%%%%%%%%%%%%%%%%%%%%%%%%%%%%%%%%%%%%%%%%%%%%%%%%%%%%%%%%%%%%%%%%%%%%%%%%%

\subsection{Sequential Monte Carlo methods}
\label{sec:smc}

We are interested in sampling from $p\left(\boldsymbol{c}_{i\cdot}^\star  \mid \boldsymbol{y}_{i\cdot\cdot}^\star, \boldsymbol{\theta} \right)$, the distribution of unobserved adherence values conditional on the observed health measures and model parameters from the training set.
For this ``smoothing'' step, we use a Sequential Monte Carlo (SMC) algorithm.

SMC algorithms are popular for conducting inference with SSMs  \cite{Doucet01, Kantas15}, in part because SMC takes advantage of the graphical structure of SSMs to perform inference.
A brief description of a standard SMC algorithm can be found in Appendix~\ref{sec:smc_intro}.
In our context, we use the particle Gibbs with ancestor sampling (PGAS) algorithm, due to Lindsten et al. \cite{Lindsten14} (building
on the particle Gibbs algorithm proposed in Andrieu et al. \cite{Andrieu10}
and backward sampling in Whiteley \cite{whiteleycomment}). The algorithm is used to draw from $p(\boldsymbol{c}_{i\cdot}^\star, \boldsymbol{\alpha}_{i\cdot\cdot}\mid \boldsymbol{y}_{i\cdot\cdot}^\star)$, the distribution of adherence and the latent health states given the observed health measures and other model parameters.
We describe below the specifics of the PGAS algorithm in detail.
Focusing on one patient, we denote the health measure observations by $y_{t}$ at time $t$, and the latent variable by $\boldsymbol{\eta}_t$ at time $t$; here, $\boldsymbol{\eta}_t$ contains $\alpha_{it}$ and $c_{it}$.
We assume the model parameters $\boldsymbol{\theta}$ to be fixed and remove them from the notation.

The algorithm runs independently for each patient, so we drop the index $i$ to simplify notation.
We assume $P$ particles progressing through time $t \in \{ 1, \dots T \}$ for an arbitrary patient.
Let $\boldsymbol{y}_{1\cdot}^\star$ be the $K$-vector denoting the health measures at time $t=1$ for the patient, with $\boldsymbol{y}_{1\cdot}^\star = (y_{11}^\star , y_{12}^\star , \cdots , y_{1K}^\star )$.
The two unobserved processes, the adherence and the latent health process, are grouped together to form the particles.
We let $\boldsymbol{\eta}_{s}^p = \{ \boldsymbol{\alpha}_{s\cdot}, c_{s}^\star \}^p$ denote the $p$-th particle for the patient at time $t$. 

During the resampling step, particles are resampled to achieve uniform weights by performing multinomial sampling on the set of possible ancestor indices.
If particle $\boldsymbol{\eta}_s^p, p \in \{1,\ldots,P\}$ is generated through a resampling step that originates from particle $\boldsymbol{\eta}_{s-1}^q, q \in \{1,\ldots,P\}$ at time $t = s-1$, then we say the index of the immediate ancestor of particle $\boldsymbol{\eta}_{s}^p$, denoted by $a_{s}^p$, is $q$.
The collection $\boldsymbol{\eta}_{1:s}^p = (\boldsymbol{\eta}_{1:s-1}^{a_{s}^p}, \boldsymbol{\eta}_{s}^p)$ is then known as the path or trajectory of particle $p$ from time $t = 1$ to $s$.

The description of the algorithm steps is modified from Lindsten et al. \cite{Lindsten14}.
The first iteration of the algorithm is a single pass of a standard SMC algorithm, which uses no reference trajectory.
At the end of this iteration, there are $P$ sampled trajectories from $t=1,\dots,T$.
A single trajectory is sampled from these $P$ trajectories using multinomial sampling with weights proportional to the likelihood of the observed data for each trajectory. 
This sampled trajectory is then used as the reference trajectory in the next iteration of the algorithm.
This selection procedure for the reference trajectory is repeated for all future iterations.

In step $t = 1$, the particles $\boldsymbol{\eta}_{1}^p$ for $p \in \{1, \dots, P-1\}$ are initialized by simulating from the distribution of the state model conditional on $\boldsymbol{\theta}$.
The last particle is set to the reference trajectory, $\boldsymbol{\eta}_{1}^P = \boldsymbol{\eta}_{1}^\prime$.
The normalized weights for each particle $w_1^p$, $p \in \{1, \dots, P\}$ are computed proportional to the likelihood of the observed data $p(\boldsymbol{y}_{1\cdot}^\star \mid \boldsymbol{\eta}_{1}^p, \boldsymbol{\theta})$.

In step $t \in \{2, \dots, T\}$, ancestor indices $a_{t}^p$ are sampled for particles $p = \{1, \dots , P-1\}$ using multinomial sampling with the weights $w_{t-1}^p$, such that $P(a_{t}^p = q) = w_{t-1}^q$.
Next, from the ancestors, the particles are transitioned to time $t$ according to the model, $p(\boldsymbol{\eta}_{t}^p \mid \boldsymbol{\eta}_{t-1}^{a_{t}^p}, \boldsymbol{\theta})$, to obtain $\boldsymbol{\eta}_t^p$ for $p = \{1, \dots , P-1\}$. 
We now transition adherence according to $p(c_{t}^\star \mid c_{t-1}^\star, \boldsymbol{\theta})$, and then transition the latent health process according to $p(\boldsymbol{\alpha}_{t\cdot} \mid \boldsymbol{\alpha}_{t-1\cdot}, c_t^\star, \boldsymbol{\theta})$.
The last particle is again set to the reference trajectory, $\boldsymbol{\eta}_{t}^P = \boldsymbol{\eta}_{t}^\prime$.

The ancestor resampling step follows, which is unique to the PGAS algorithm.
A new ancestor index $a_{t}^{P}$ is sampled for the reference trajectory $\boldsymbol{\eta}_{t}^P$ using multinomial sampling with normalized weights $\tilde{w}_{t-1}^p$ proportional to the state transition model density $p(\boldsymbol{\eta}_{t}^p \mid \boldsymbol{\eta}_{t-1}^{a_{t-1}^p}, \boldsymbol{\theta})$ for $p \in \{1, \dots, P\}$. To complete the step, the trajectories are updated.
We set $\boldsymbol{\eta}^p_{1:t} = (\boldsymbol{\eta}_{1:t-1}^{a_{t}^p}, \boldsymbol{\eta}_t^p)$ for all $p \in \{1, \dots, P\}$.
The weights $w_t^p$ for $p \in \{1, \dots, P\}$ are computed, ready for the next step of the algorithm.

In the final step $t = T$, a particle index $p$ is sampled using the weights $w_{T}^p$. This step completes one iteration of the algorithm, and the sampled trajectory $\boldsymbol{\eta}^p_{1:T}$ is returned as the output.
The next iteration of the algorithm uses this sampled trajectory as the reference trajectory.
We run $g$ iterations for each of the $h$ draws of $\boldsymbol{\theta}$, and a proportion of the initial iterations are discarded as burn-in.
We choose a burn-in proportion of 20\%, so for each patient we retain $4/5 g \times h$ draws of the path from time $1$ to $T$.
The techniques in Jacob et al. \cite{jacob2019smoothing} could be used to remove the burn-in bias and provide unbiased estimators of smoothing quantities.

%%%%%%%%%%%%%%%%%%%%%%%%%%%%%%%%%%%%%%%%%%%%%%%%%%%%%%%%%%%%%%%%%%%%%%%%%%%%%%%%%%%%%%%%%%
% Application to hypertensive cohort
%%%%%%%%%%%%%%%%%%%%%%%%%%%%%%%%%%%%%%%%%%%%%%%%%%%%%%%%%%%%%%%%%%%%%%%%%%%%%%%%%%%%%%%%%%

\section{Application to antihypertensive medication adherence}
\label{sec:hyp}

Hypertension, or high blood pressure, is a widespread and serious health problem for which effective treatment is available, but currently underutilized.
Hypertension can cause serious complications, including heart disease, stroke, renal disease, and a shorter life expectancy \cite{Amery85, Shep91}.
Almost 33\% of American adults have hypertension, while for African American adults, the rate is 44\%, which is among the highest prevalence of hypertension in the world \cite{Go13}.
Because of the widespread prevalence of hypertension, the accessibility of effective treatments, and the under-utilization of treatment, Farley et al. \cite{Farley10} suggested that additional treatment for hypertension would be the most effective clinical preventive service for reducing preventable deaths.
They predicted that every 10\% increase in hypertension treatment would prevent an additional 14,000 non-elderly deaths per year.

Hypertension is often called ``a silent killer'' because most patients do not experience any noticeable symptoms, but can have severe adverse health events due to uncontrolled hypertension.
Low medication adherence to antihypertensive drugs is a widespread problem, and has been shown to contribute to poor blood pressure control \cite{Zeller08, Yue15}.
Although over 75\% of adults with hypertension are taking antihypertensive medication, only about half of adults with hypertension have their blood pressure controlled in a healthy range \cite{Nwankwo13, Go13}.
Higher medication adherence to antihypertensive drugs has been shown to improve patient outcomes.
Corrao et al. \cite{Corrao11} found that taking 50-75\% of prescribed doses resulted in a reduced risk of cardiovascular outcomes of 20\%, and taking at least 75\% of doses resulted in a reduced risk of 25\%.

Multiple studies have found that patients struggle with adherence to antihypertensive medication.
The lack of noticeable symptoms is likely one reason for low adherence rates, as patients do not experience any immediate effects of being lax in their adherence.
Hypertension is also a chronic condition, so patients often are prescribed medication for months or years, and must continue being vigilant to prevent adverse health events \cite{KrouselWood04}.
Vrijens et al. \cite{Vrijens08} found that about half of patients stopped taking their medications within one year.
Even among patients who persist, many do not take the recommended number of doses.
Corrao et al. \cite{Corrao11} found that of those patients who persisted to take their medications, about half had adherence less than 50\%.

%%%%%%%%%%%%%%%%%%%%%%%%%%%%%%%%%%%%%%%%%%%%%%%%%%%%%%%%%%%%%%%%%%%%%%%%%%%%%%%%%%%%%%%%%%
% Data description
%%%%%%%%%%%%%%%%%%%%%%%%%%%%%%%%%%%%%%%%%%%%%%%%%%%%%%%%%%%%%%%%%%%%%%%%%%%%%%%%%%%%%%%%%%

\subsection{Data description}
\label{sec:hyp_data}

Our data are from the pre-randomization baseline period of a clinical trial that tested the effect of an intervention to increase provider-patient communication skills on hypertensive patient outcomes \cite{Kressin15}.
The outcomes of interest in the trial included both adherence to antihypertensive medication and the level of blood pressure control. 

Participants were enrolled in the study during the period between August 2004 and June 2006.
Recruitment was done at seven outpatient primary care clinics at the Boston Medical Center, an inner-city safety-net hospital affiliated with the Boston University School of Medicine.
Patients had to meet certain eligibility criteria, including that they were of white or black (African or Caribbean) ethnicity, 21 years of age or older, had an outpatient diagnosis of hypertension on at least 3 occasions, and must have been currently taking antihypertensive medication at the time of enrollment.
Recruitment resulted in an initial cohort of 869 patients.
For the statistical analysis, only anonymized patient data was used to maintain patient privacy. 

Detailed medication adherence was measured using the Medication Event Monitoring System (MEMS), developed by Aardex Group, Ltd, Sion, Switzerland.
A MEMS cap is an electronic cap for medication bottles that records the date and time of each bottle opening.
Each patient was given one MEMS cap and was instructed to use the MEMS cap with their most frequently taken antihypertensive medication.
The MEMS caps were collected from patients at the end of the study, allowing researchers to download the timings of all bottle openings.
Each patient had a different number of days on which medication adherence was recorded, depending on enrollment time and when the MEMS cap was returned.

A patient was assumed to be adherent to their medication regime if the number of times they opened the bottle on a particular day matched the prescribed dosing frequency, and non-adherent otherwise.
Blood pressure was measured during routine clinical care, resulting in irregularly-spaced observations of blood pressure measurements.
Our analyses were restricted to the subset of patients who returned the MEMS cap and had at least one blood pressure reading.
Patients missing one or both of these components provide no information about the relationship between their adherence and blood pressure, so they would not be informative to include the analysis.
This inclusion criteria resulted in an analysis on 503 patients out of the initial cohort of 869 patients.

Socio-demographic and health covariates were also collected from the study participants.
Socio-demographic data included race, gender, and age at the start of the study.
Baseline comorbidities that might have influenced blood pressure levels were collected from electronic health records.
These included presence/absence of cerebrovascular disease, congestive heart failure, chronic kidney disease, nicotine dependence, coronary artery disease, diabetes mellitus, hyperlipidemia, peripheral vascular disease, benign prostatic hypertrophy, and obesity (BMI greater than 30 $kg/m^2$).

\begin{figure}
\centering
\includegraphics[width = 5in]{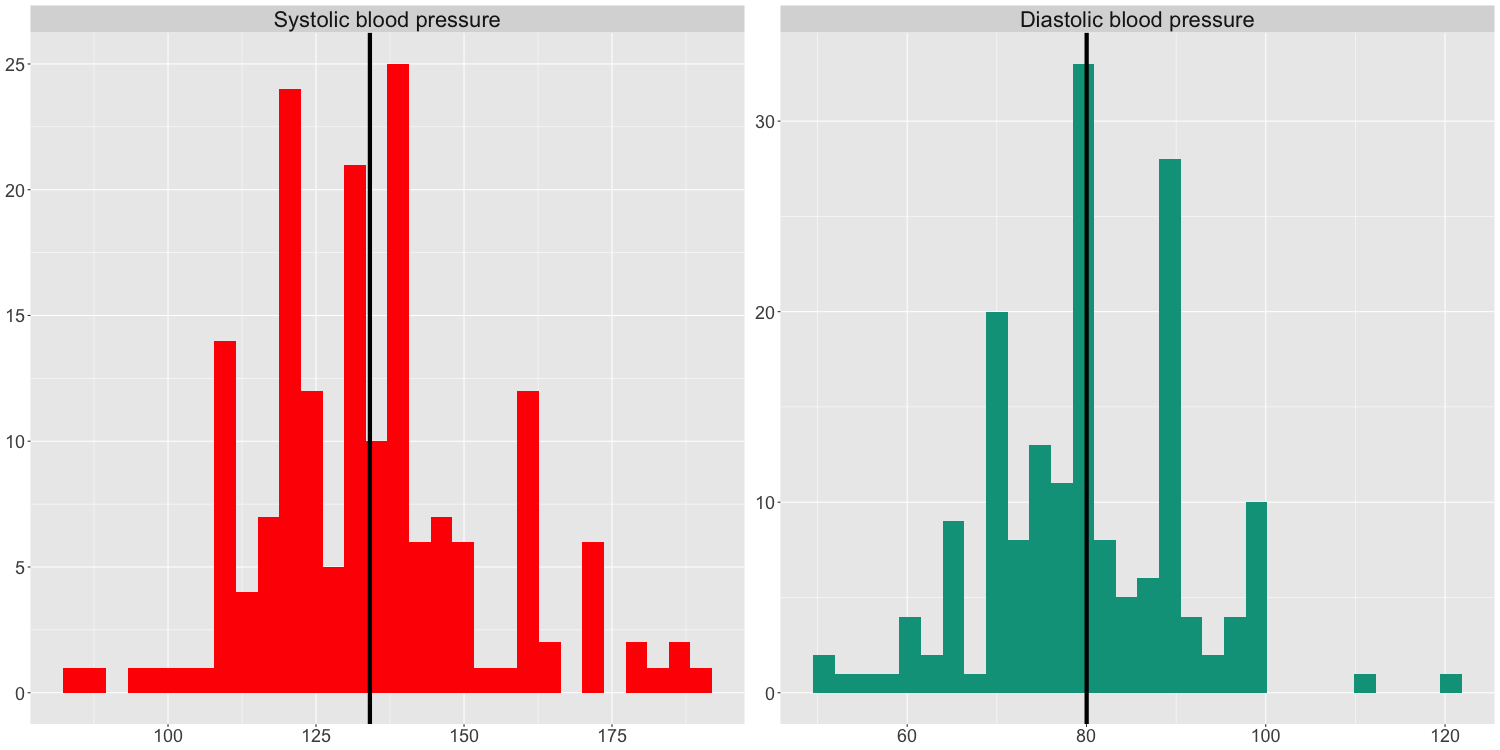}
\caption{Distribution of initial blood pressure\label{fig:bp}}
\end{figure}

\begin{figure}
\centering
\includegraphics[width = 5in]{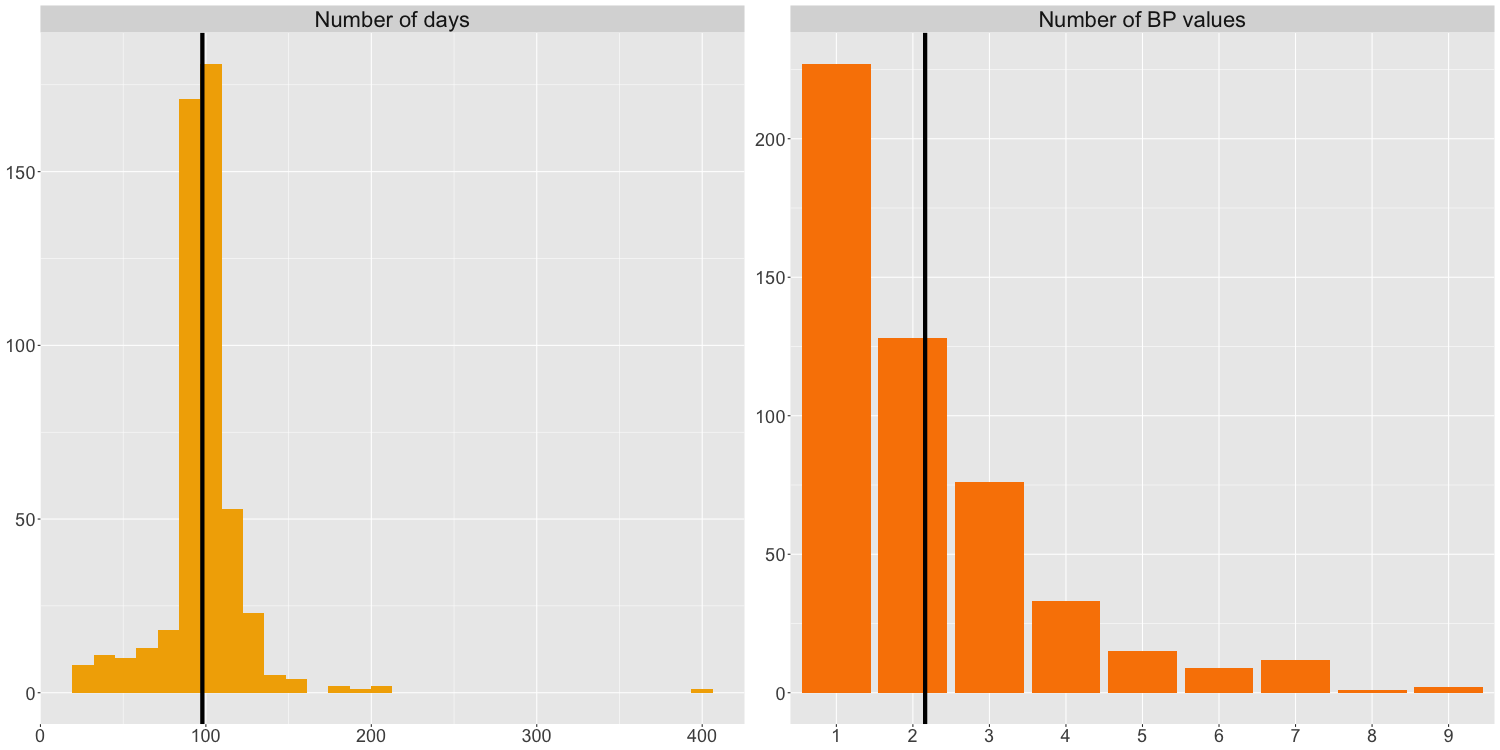}
\caption{Distribution of observed counts of adherence and blood pressure\label{fig:counts}}
\end{figure}

\begin{figure}
\centering
\includegraphics[width = 3in]{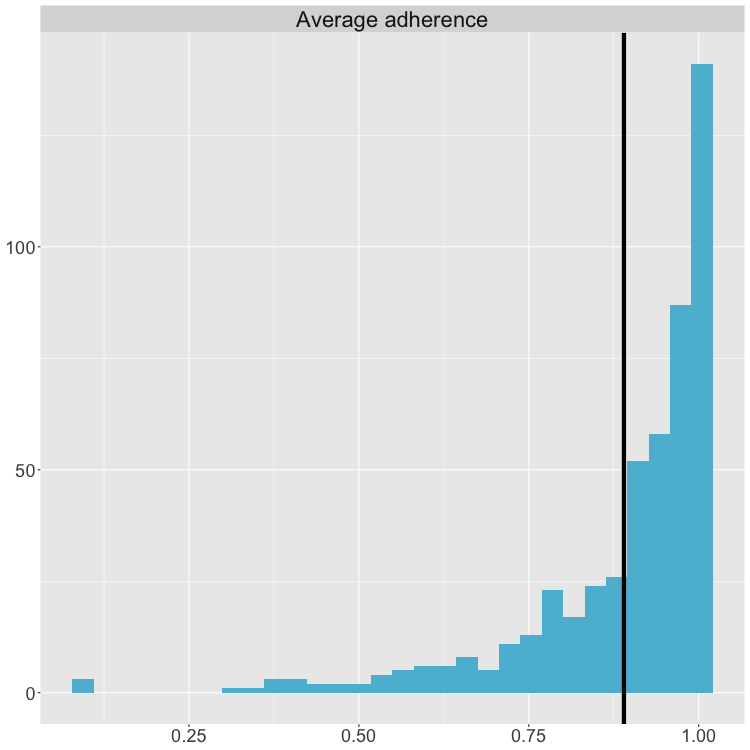}
\caption{Distribution of average adherence\label{fig:ad}}
\end{figure}

Figures \ref{fig:bp} - \ref{fig:ad} show the distributions of key variables, where the vertical black line on each figure indicates the mean of the distribution.
The mean diastolic blood pressure at the start of the study was $80$ mm Hg (Figure~\ref{fig:bp}), which is the upper end of the recommended range for diastolic blood pressure \cite{Whelton18}.
The mean systolic blood pressure at the start of the study was $134$ mm Hg, which is above the recommended range. 
The mean number of days for which adherence was observed was 98 days (Figure~\ref{fig:counts}).
In contrast, the mean number of observed blood pressure values was slightly above 2, with many patients having only 1 or 2 measurements.
Finally, the mean value for mean daily adherence over the observed time window for each patient was quite high, at $88\%$ (Figure~\ref{fig:ad}).

The covariates, which are all binary, are summarized in Table \ref{tab:bin_cov}.
The cohort was mostly female ($68\%$), and majority black ($54\%$).
The mean age was 60, and most people fell in the age range of $51-70$ (62\%).
Obesity was the most common comorbidity ($60\%$), followed closely by hyperlipidemia ($57\%$).
Diabetes and coronary artery disease were also relatively common ($36\%$ and $15\%$, respectively).
Benign prostatic hypertrophy, renal insufficiency, nicotine dependence, 
peripheral vascular disease and cerebral vascular disease all had low incidence rates (3\% to 7\%).

In the antihypertensive cohort, some patients had missing adherence indicators throughout the study period.
There were $70$ patients with at least one day of missing adherence, but most of these patients had only a small amount of missingness; $50$ of these $70$ patients had only one or two missing values.
Adherence may have been missing due to MEMS cap malfunctions, hospital inpatient stays, or other reasons.
We assume that the missing adherence indicators are missing at random,
so when fitting the baseline adherence model we exclude missing days.
For the health measures state-space model in which adherence is a predictor, missing adherence can be modeled explicitly.
Details involving inference with missing adherence information were described in Campos et al. \cite{Campos20}.

\begin{table}
\centering
\begin{tabular}{l l}
Variable				        & Percent	\\ \hline
Female							& 68 		\\
Black							& 54 		\\
Age 50 and below                & 17        \\
Age 51 - 60		                & 29 		\\
Age 61 - 70			            & 33 		\\
Age above 70					& 21 		\\
Obese							& 60 		\\
Hyperlipidemia					& 57 		\\
Diabetes						& 36 		\\
Coronary artery disease			& 15		\\
Renal insufficiency				& 7			\\
Nicotine dependence				& 6 		\\
Peripheral vascular disease		& 6			\\
Cerebral vascular disease		& 5			\\
Benign prostatic hypertrophy    & 3         \\
\end{tabular}
\caption{Summary of binary covariates}
\label{tab:bin_cov}
\end{table}

%%%%%%%%%%%%%%%%%%%%%%%%%%%%%%%%%%%%%%%%%%%%%%%%%%%%%%%%%%%%%%%%%%%%%%%%%%%%%%%%%%%%%%%%%%
% Application to hypertensive cohort
%%%%%%%%%%%%%%%%%%%%%%%%%%%%%%%%%%%%%%%%%%%%%%%%%%%%%%%%%%%%%%%%%%%%%%%%%%%%%%%%%%%%%%%%%%

\subsection{Inference for unobserved adherence}
\label{sec:hyp_application}

To evaluate the performance of our approach, we divided the hypertensive cohort at random into a training and test sample.
The $503$ patients in the cohort were split into a training set of $n_{train} = 400$ patients and a test set of $n_{test} = 103$ patients.
We first fit our state-space and logistic regression model on the training sample using the approach in Section~\ref{sec:model}, with $M = 20$ imputations of missing adherence values, to obtain the posterior samples from $p(\boldsymbol{\theta}\mid \boldsymbol{c}^\prime, \boldsymbol{y}^\prime)$.
For our application, outcomes were the bivariate blood pressure measures ($K=2$) of diastolic and systolic blood pressures.

We then applied our adherence estimation algorithm from Section~\ref{sec:smc} to the test set, holding out the adherence information as if it were unobserved.
The predictions are summarized using intervals for average adherence for each patient, where the average is computed over the time window during which adherence was actually observed for each patient, 
that is, $\bar{a}_i = \sum_{t=1}^{T_i}c_{it}/T_i$.
We then compared the predictive intervals for each patient to the actual adherence recorded in the test set to evaluate performance.
Previous work has routinely used aggregate measures of adherence such as average adherence or adherence categories, so we follow this precedent \cite{Wu08, KrouselWood04, Balkrishnan03, Shalansky04}.
However, the procedure produces full vectors of possible adherence for each patient, so more detailed summaries could also be investigated if they are of interest to a researcher, including time trends or measures of variation in adherence.

%%%%%%%%%%%%%%%%%%%%%%%%%%%%%%%%%%%%%%%%%%%%%%%%%%%%%%%%%%%%%%%%%%%%%%%%%%%%%%%%%%%%%%%%%%
% Results
%%%%%%%%%%%%%%%%%%%%%%%%%%%%%%%%%%%%%%%%%%%%%%%%%%%%%%%%%%%%%%%%%%%%%%%%%%%%%%%%%%%%%%%%%%

\subsubsection{Training data model summaries}
\label{sec:posterior}

We chose prior distributions for the adherence and health measures state space model based on the context.
For the adherence logistic regression model, we take an approach recommended by Gelman \cite{Gelman06}. 
We assume an improper uniform prior on $\boldsymbol{\lambda}$, and an improper uniform prior on $(0, A)$ as $A \rightarrow \infty$ for $\sigma_{\delta}$.
For the blood pressure state-space model, we place uninformative but still restricted priors for systolic, $k = 1$, and diastolic, $k = 2$, independently:
\begin{align*}
\rho_{k} & \sim U(-1,1)\\
\phi_{k} & \sim N(0,25)\\
\gamma_k &\sim N(0, 25)\\
\sigma_{\varepsilon k} &\sim U(0, 30), \rho_\varepsilon  \sim U(-1,1)\\
\sigma_{\nu k} &\sim U(0, 10), \sigma_{0 k} \sim U(0, 30)\\
\beta_{11}&\sim N(120, 400), \beta_{12}\sim N(80, 400)\\
\beta_{jk}&\sim N(0, 400), j = 2, ..., p\\
\sigma_{U k} &\sim U(0, 30).
\end{align*}

The model fitting on the training data that resulted in posterior draws from $p(\boldsymbol{\theta} \mid \boldsymbol{c}^\prime, \boldsymbol{y}^\prime)$ was performed separately for the adherence data and for the blood pressure measures given the adherence data.
We fit our adherence and blood pressure models via STAN \cite{Carpenter17} to obtain posterior draws of the model parameters.
For both the adherence and blood pressure models, the algorithm was run with four chains for 80,000 iterations, which included a burn-in period of 40,000 iterations, which is the default burn-in ratio in STAN.
Trace plots and the Gelman-Rubin convergence statistic \cite{Gelman92} suggested the chains had converged.
 
The posterior means and standard deviations of the parameters for the adherence data logistic regression model are summarized in Table~\ref{tab:theta_a}.
The posterior distribution of the standard deviation of the random effect is centered around 1.7, resulting in most posterior draws of $\alpha$ for patients in the test set being between $-5$ and $5$, demonstrating that patient adherence had a fair amount of variability.
A full summary of the distribution of model parameters of the blood pressure state-space model can be found in Campos et al. \cite{Campos20}.

\begin{table}
\centering
\label{tab:theta_a}
\begin{tabular}{l l l l}
            &       & 2.5\%     & 97.5\%\\
Parameter   & Mean  & Quantile  & Quantile \\ \hline
Intercept & 2.11 & 1.54 & 2.67\\
Male & 0.33 & -0.10 & 0.77\\
Age 51 - 60 & 0.40 & -0.15 & 0.96\\
Age 61 - 70 & 0.66 & 0.10 & 1.23\\
Age > 70 & 1.10 & 0.46 & 1.73\\
White & 0.56 & 0.16 & 0.96\\
Obese & 0.01 & -0.38 & 0.40\\
Nicotene dependence & -0.23 & -0.96 & 0.52\\
Hyperlipidemia & 0.23 & -0.17 & 0.63\\
Diabetes & -0.32 & -0.73 & 0.10\\
Peripheral vascular disease & -0.05 & -0.87 & 0.77\\
Renal insufficiency & 0.12 & -0.67 & 0.91\\
Benign prostatic hypertrophy & -0.33 & -1.41 & 0.76\\
Coronary artery disease & -0.06 & -0.62 & 0.50\\
Congestive heart failure & -0.35 & -1.38 & 0.68\\
Cerebral vascular disease & 0.28 & -0.54 & 1.10\\
Random effect standard deviation & 1.74 & 1.58 & 1.91\\
Alpha & 0.00 & -3.41 & 3.41\\
\end{tabular}
\caption{Summary of marginal posterior distributions of adherence model parameters}
\end{table}

\subsubsection{Estimated adherence results}
\label{sec:estimation}

To estimate the average adherence per patient for the test sample, we applied PGAS as described in Section~\ref{sec:smc}.
For each patient, we used $P = 32$ particles with $e = 100$ iterations and a burn-in period of $20$ steps.
The SMC algorithm was run for $f = 100$ different draws from the posterior distribution $p(\boldsymbol{\theta} \mid \boldsymbol{y}^\prime, \boldsymbol{c}^\prime)$, resulting in $e \times f = 10,000$ draws.

Average adherence for each patient was summarized using credible intervals calculated from the empirical distribution across the particles.
For each imputation, on 32 cores with 4000 MB per CPU with Intel `Broadwell' core types, the computation for all 103 patients in the test set took 30 minutes.

\begin{figure}
\centering
  \includegraphics[width = 3in]{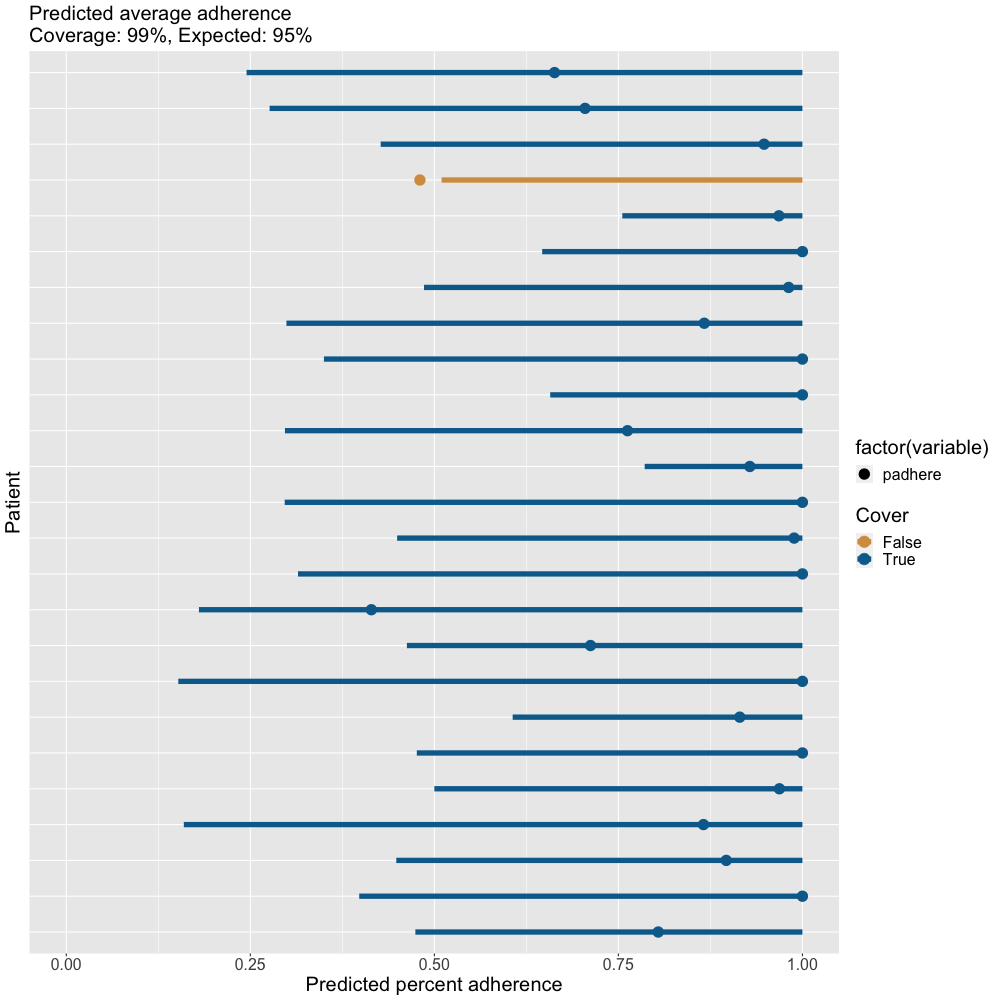}
\caption{\label{fig:coverage_95} Estimated average adherence: 95\% intervals}
\end{figure}

\begin{figure}
\centering
  \includegraphics[width = 3in]{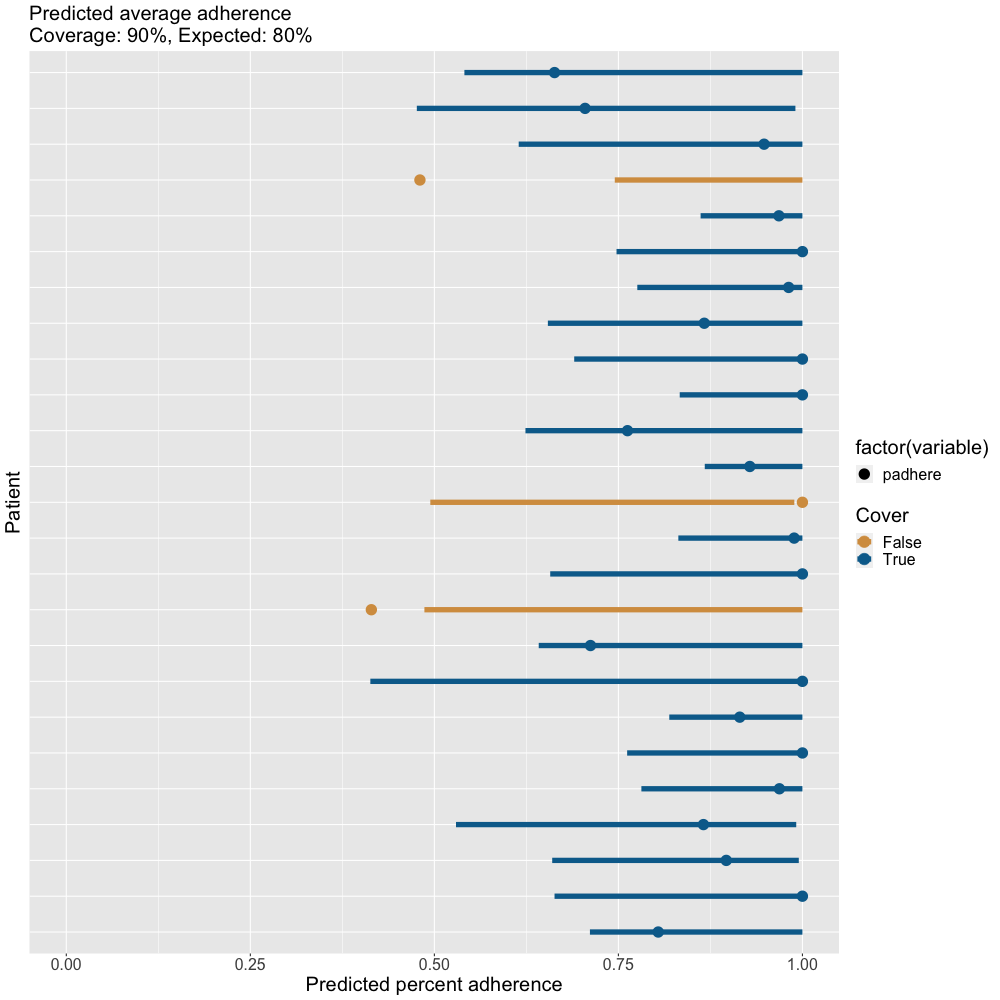}
\caption{\label{fig:coverage_80} Estimated average adherence: 80\% intervals}
\end{figure}

\begin{figure}
\centering
  \includegraphics[width = 3in]{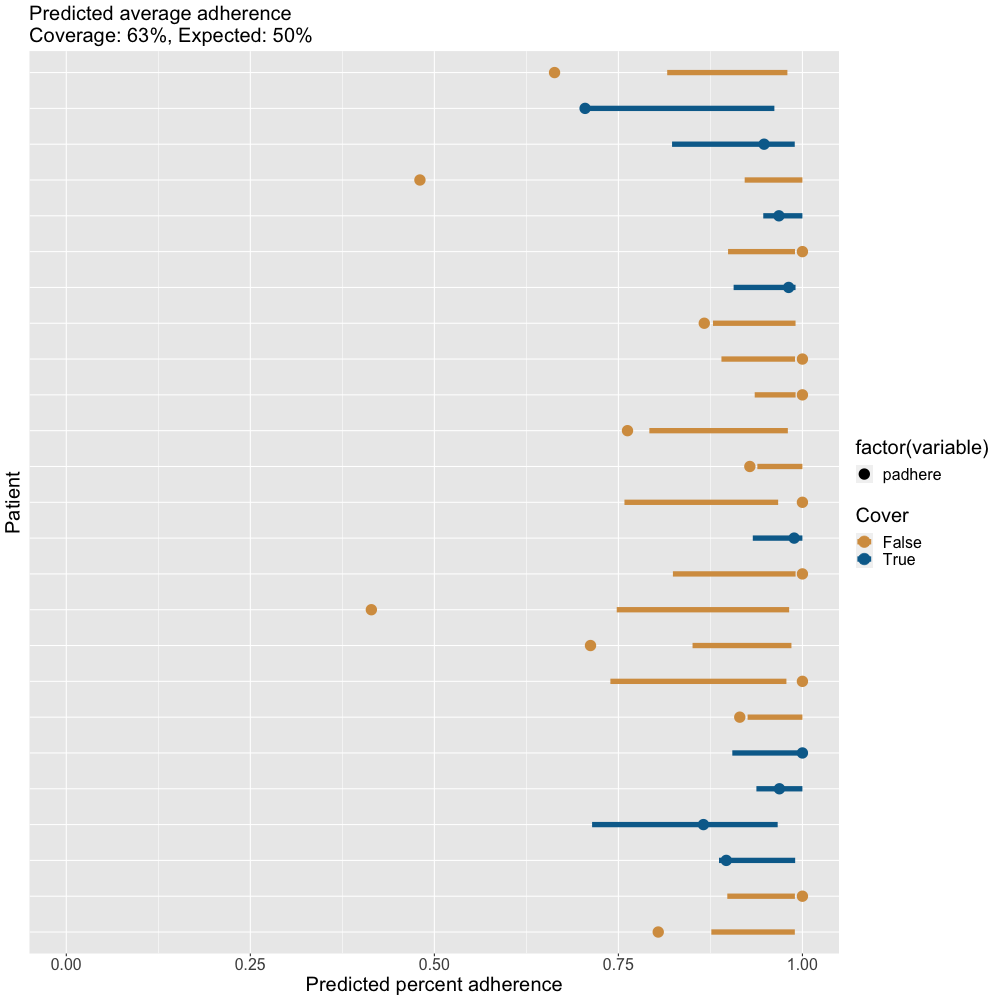}
\caption{\label{fig:coverage_50} Estimated average adherence: 50\% intervals}
\end{figure}

We constructed three sets of central posterior interval summaries for average adherence for the 103 patients corresponding to 95\%, 80\% and 50\% (Figures~\ref{fig:coverage_95}-\ref{fig:coverage_50}).
These figures show intervals for 25 randomly selected patients, but the coverage rate was calculated for all $103$ test patients.
The circular dots are the true value of average adherence for each patient.
Intervals that are dark blue cover the true average adherence for those patients, and intervals that are gold do not cover the true average adherence.
The square dots show the estimated average adherence from the logistic regression model that is based on only baseline covariates and excludes blood pressure, which we call the baseline covariate model.

To check convergence of the algorithm, first we repeated the algorithm 10 times with different starting seeds.
Coverage was quite stable across runs. For example, coverage of the 80\% intervals ranged from 89\% to 90\%.
Next, we also repeated the algorithm using combinations of a larger number of particles and iterations to see if results changed (up to $P =128$ particles and $e = 4000$ iterations).
Coverage did not change with increased particles or iterations; for 80\% intervals, coverage ranged from 89\% to 91\%.
The stability of the results implied that we used a sufficient number of particles and number of iterations to reach convergence.

\begin{figure}
\centering
  \includegraphics[width = 2.5in]{coverage_80.png}
  \includegraphics[width = 2.5in]{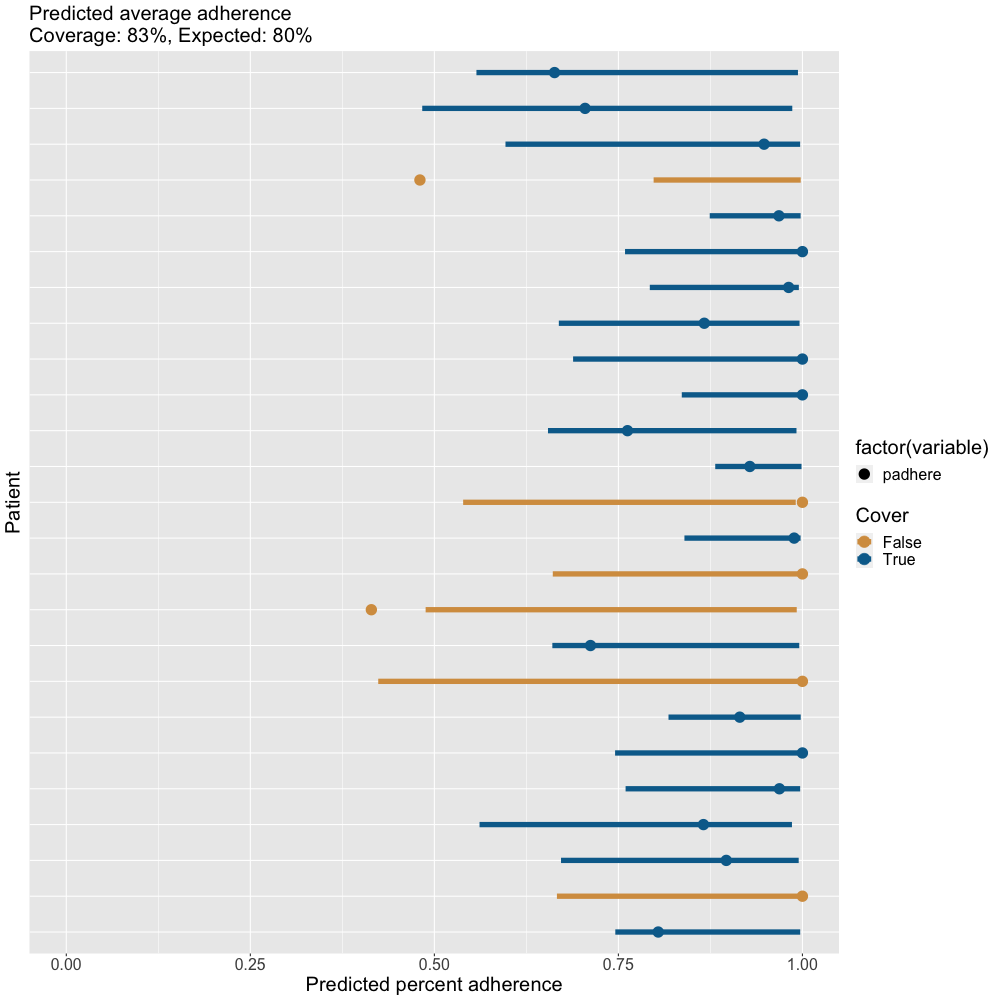}
\caption{\label{fig:coverage_ad_80} Estimated average adherence 80\% intervals. Blood pressure is included in the left plot, while the right plot only considers baseline variables.}
\end{figure}

Overall, the intervals all demonstrate some overcoverage (Table~\ref{tab:intervals}).
For the following discussion, we focus on the 80\% intervals because they strike a balance between coverage and interval width. 
We further investigated some of the patterns shown in the results.
We compared the algorithm with blood pressure information to the baseline covariate model.
In the baseline covariate model, posterior draws were generated from the logistic regression of average adherence on baseline covariates, including comorbidities and socio-demographic information (Figure~\ref{fig:coverage_ad_80}).
Taking into account blood pressure results in slightly wider intervals: the mean interval width without blood pressure is $0.32$, and the mean interval width with blood pressure is $0.33$.
In addition, the coverage increases from 83\% to 90\%, so the intervals with blood pressure are farther from nominal coverage.

\begin{table}
\centering
\label{tab:intervals}
\begin{tabular}{r r r r}
Expected & Actual & Mean & Max \\
coverage & coverage & interval width & interval width \\ \hline
0.95 & 0.99 & 0.61 & 0.94\\
0.80 & 0.90 & 0.34 & 0.68\\
0.50 & 0.63 & 0.14 & 0.31\\
\end{tabular}
\caption{Summary of posterior intervals}
\end{table}

We next consider factors that may impact interval width.
The most striking pattern is a very high negative correlation between interval length and estimated average adherence using the baseline covariate model -- patients with a high estimated average adherence from the baseline covariate model have the narrowest intervals.
This pattern could be due to the very high level of adherence in this population.
For patients with possibly low predicted adherence rates, there are fewer training cases, possibly resulting in wider intervals.
This pattern implies model performance could be different on a population with less homogeneous adherence behavior.

Interval width also varied, although more moderately, with other variables.
Patients who were older, had congestive heart failure, had peripheral vascular disease, had renal insufficiency, or had nicotene dependence tend to have longer intervals (listed in decreasing correlation, with age having a correlation around $0.2$ with interval width, ranging down to around $0.1$ for nicotene dependence).
Patients who had diabetes or had benign prostatic hypertrophy have narrower intervals (around $0.1$ negative correlation).

%%%%%%%%%%%%%%%%%%%%%%%%%%%%%%%%%%%%%%%%%%%%%%%%%%%%%%%%%%%%%%%%%%%%%%%%%%%%%%%%%%%%%%%%%%
% Discussion
%%%%%%%%%%%%%%%%%%%%%%%%%%%%%%%%%%%%%%%%%%%%%%%%%%%%%%%%%%%%%%%%%%%%%%%%%%%%%%%%%%%%%%%%%%

\section{Discussion}
\label{sec:discussion}

In this paper, we have introduced a method for inferring a patient's medication adherence.
Our method estimates adherence for patients with no observed adherence behavior, and incorporates both fixed characteristics and time-varying health measures.
We use a state-space model to flexibly incorporate the time series structure and irregularly observed health measures.
To achieve efficient computation, we use a Sequential Monte Carlo algorithm to conduct inference on adherence behavior.
The procedure involves fitting a model on a training set of patients with detailed adherence information available, such as data collected from electronic monitoring medication bottle caps.
Once the model is trained on the population of interest, predictions are made for new patients using only commonly-collected clinical data, including health information and demographic covariates.
Thus, this procedure is possibly more resource-intensive in the first phase than other methods of estimating adherence, but once the model has been trained, no specialized or additional information needs to be collected, and thus can be easily scaled to a large population.
The estimated average adherence intervals produced by our method could be useful in clinical practice as an alternative to patient self-reporting or tracking prescription refills.

The method is a two-step procedure, in which inference on model parameters is conducted on a training set, and adherence values are inferred on a test set.
In conducting inference using a two-step procedure, we depart from a fully Bayesian approach, which would perform inference on all unknown parameters jointly over both the training and test set.
Instead, we chose to perform inference separately on the training and test set with two purposes in mind.
First, this approach permits us to easily check out-of-sample predictability.
Second, and more importantly, our two-step approach more closely mirrors what would occur if such a method were applied in clinical practice.
In a clinical setting, a set of posterior draws of the model parameters would be generated once using a large set of training data.
Then, each new patient would have their medication adherence inferred given this fixed set of posterior model parameter draws.
Once inference on model parameters is complete on the training set, inference for a new patient can be computed quickly; for example, in our application predictions for a single patient took less than a minute, and with further computational optimization this time could likely be reduced.

One possible limitation to our approach is that some patients could be falsely flagged as being low-adherers when in fact their blood pressure is fluctuating because they are not responding well to their treatment plan.
However, if a patient is showing problematic health trends that are similar to patterns seen in low-adherers, it is a positive benefit that their healthcare provider be notified, regardless of the underlying cause.
The goal of our predictions is to better inform provider-patient discussions, and such information might result in a conversation between provider and patient as to the treatment plan.

Our procedure is flexible and can easily be adapted to a wide variety of medical contexts in which medication adherence is of interest.
The modularity of the framework permits many different modeling choices.
The baseline covariate adherence models could be extended beyond random effects logistic regression, and predictions from such models would easily flow into the rest of the algorithm.
Alternatively, we can incorporate different modeling assumptions to the state-space model.
For example, although we assume Normality for many of the distributions, our approach
could accommodate non-Normal data, though the PGAS algorithm relies on being able to evaluate the transition density, so some computational details would have to be adapted.
In the case of a transition density that is difficult to evaluate, the user could substitute an alternative SMC algorithm.
In considering other contexts, the procedure would likely be particularly useful in settings with regularly observed health measures.
For example, patients can now easily monitor certain health measures at home using technology such as ambulatory blood pressure monitors and smartphone apps.
Richer outcome measures would better inform the model and could provide tighter intervals for patient adherence.

In order to apply this procedure in practice, further research would need to be done to evaluate the performance in different settings.
For example, with a population with more variability or lower adherence, the model might require more training data in order to achieve accurate and stable results.
Further investigation would be necessary to determine how large and diverse of a sample is necessary.
Exploration into whether different institutions or patient populations would result in substantially different models for the same health measures would also be of both scientific and practical interest.

Inferred medication adherence behavior from our approach could be used in a clinical decision tool for providers. 
For example, our approach could form the basis of the development of a mobile phone app used by providers during clinic visits with their patients.
The provider would import relevant information from electronic health records, such as baseline characteristics and health measurements over time, and enter (or import) the relevant health measures (blood pressure measures, in the context of a decision tool for antihypertensive medication
adherence) along with corresponding dates of clinic visits.
The app would then provide interval estimates of the patient's medication adherence.
A tool that estimates recent adherence behavior would empower patients and providers to have more informed discussions about medication adherence and treatment plans.

\section*{Supplementary Material}
We have shared the analysis code, including an application to simulated data, in a GitHub repository at \url{https://github.com/lfcampos/Medication-Adherence}.

\section*{Acknowledgements}
We greatly thank Pierre E. Jacob for his substantial contribution and feedback on this work.
We thank the Health Statistics Working Group at Harvard University for their discussions and comments.

This work was supported by grants from the Agency for Healthcare  Research and Quality (AHRQ grant R03-HS022112) and the National Heart, Blood, and Lung Institute (NHLBI grant R21-HL121366).
Kristen Hunter was supported by the Department of Defense (DoD) through the National Defense Science \& Engineering Graduate Fellowship (NDSEG) Program.
Luis Campos would like to acknowledge that this research is in no way related to his work at Etsy, Inc.

%%%%%%%%%%%%%%%%%%%%%%%%%%%%%%%%%%%%%%%%%%%%%%%%%%%%%%%%%%%%%%%%%%%%%%%%%%%%%%%%%%%%%%%%%%
% References
%%%%%%%%%%%%%%%%%%%%%%%%%%%%%%%%%%%%%%%%%%%%%%%%%%%%%%%%%%%%%%%%%%%%%%%%%%%%%%%%%%%%%%%%%%

\bibliographystyle{plainnat}

%%%%%%%%%%%%%%%%%%%%%%%%%%%%%%%%%%%%%%%%%%%%%%%%%%%%%%%%%%%%%%%%%%%%%%%%%%%%%%%%%%%%%%%%%%
% Appendices
%%%%%%%%%%%%%%%%%%%%%%%%%%%%%%%%%%%%%%%%%%%%%%%%%%%%%%%%%%%%%%%%%%%%%%%%%%%%%%%%%%%%%%%%%%

\appendix
\section{Appendix: SMC algorithm description}
\label{sec:smc_intro}

SMC algorithms use particles $(\boldsymbol{\eta}_{1:t}^p)_{p=1}^P$, which are a finite set of time-indexed vectors from $t=1,\ldots,T$, to recursively approximate target probability distributions $p(\boldsymbol{\eta}_1,\ldots,\boldsymbol{\eta}_t|y_{1},\ldots,y_t)$.
A total of $P$ particles are initialized by drawing from the initial distribution of the states at time $t=1$: $\boldsymbol{\eta}^p_1\sim p(d\boldsymbol{\eta}_1)$.
Then, at each time point $t = s$, the weight $w_s^p$ of each particle is calculated according to the likelihood given the observed data $y_s$ at the current time point: $w_s^p =  p(y_s|\boldsymbol{\eta}_s^p)$.
The particles are then resampled with replacement with probabilities proportional to the weights so that the resampled particles have equal weights.
Resampling with replacement can be understood as assigning a number of offspring to each particle, such that the numbers of offspring per particle follow a multinomial distribution with weights proportional to $(w_s^p)_{p=1}^P$, and each offspring particle has weight $1/P$.
Thus, the sampling step induces an ancestor-offspring relationship among particles. Various resampling schemes can be implemented, depending on hardware, computational budget and desired precision \cite{douc2005comparison,murray2016parallel,gerber2019negative}.
Finally, new particles are simulated for time $t = s+1$ conditional on the resampled particles at time $t = s$ according to the state transition model, $p(\boldsymbol{\eta}_{s+1}^p|\boldsymbol{\eta}_{s}^p)$.

The process of calculating weights, resampling, and then simulating parameter vectors at the next time point is repeated until the procedure has progressed through all of the time points.
The final paths $(\boldsymbol{\eta}_{1:T})_{p=1}^P$ provide an approximation of the smoothing distribution $p(\boldsymbol{\eta}_{1:T}|y_{1:T})$ as $P$ goes to infinity.
Memory-wise, a naive implementation would store all the generated
paths for a cost of $P\times T$; however, efficient implementations can be closer to $P+T$ \cite{jacob2015path,koskela2018asymptotic}.
These implementations exploit a defect of the particle method, often called ``path degeneracy'', whereby the paths share a high amount of common ancestry, making the effective number of samples approximating the first marginals of $p(\boldsymbol{\eta}_{1:T}|y_{1:T})$ very low.
Fighting path degeneracy is the topic of a rich literature on particle smoothing, see Kantas et al. \cite[Section 4]{Kantas15} for a recent survey.
Here we will use a method called particle Gibbs with ancestor sampling (PGAS) due to Lindsten et al. \cite{Lindsten14} (building
on the particle Gibbs algorithm proposed in Andrieu et al. \cite{Andrieu10}
and backward sampling in Whiteley \cite{whiteleycomment}).
In this method the above SMC algorithm will essentially constitute the basis for one iteration, and then more iterations will be performed as in an MCMC algorithm.
After discarding an initial ``burn-in'' period we will form our smoothing estimates by averaging the path approximations over iterations.

\end{document}